\documentclass[12pt]{article}
\usepackage{graphicx}
\usepackage{float}
\input epsf.tex
\newcommand{\sn}[1]{\mbox{\textsc{#1}}} 
\def\basin{\sn{BASIN}$~$}
\def\basinn{\sn{BASIN}}
\def\cpp{C++$~$}
\def\pyth{{Python}$~$}

\def\attr{\sn{Attribute}$~$}
\def\attrs{\sn{Attribute}s$~$}
\def\blist{\sn{List}}
\def\bgrid{\sn{Grid}}
\def\data{\sn{Data}}
\def\region{\sn{Region}$~$}
\def\regions{\sn{Region}s}
\def\c{C$~$}
\begin{document}
\renewcommand{\thepage}{\roman{page}}
\title{Beowulf Analysis Symbolic INterface
BASIN: Interactive Parallel Data Analysis for Everyone\footnote{\tt http://www.physics.drexel.edu/BASIN}}
\author{E. Vesperini$^1$, D.M. Goldberg$^1$, S. McMillan$^1$, J. Dura$^2$, D. Jones$^2$ \\
\small{ $^1$ Department of Physics, Drexel University, Philadelphia, PA} \\
\small{ $^1$ vesperin@physics.drexel.edu, goldberg@drexel.edu, steve@physics.drexel.edu} \\
\small{ $^2$ Department of Computer Science, Drexel University, Philadelphia, PA} \\
\small{ $^2$ jd992@drexel.edu, dfj23@drexel.edu} 
}

\newpage

\renewcommand{\thepage}{\arabic{page}}
\setcounter{page}{1}

\maketitle
\begin{center} Submitted for publication to {\it Computing in Science and Engineering}\\special issue on Computational Astrophysics \end{center}

\begin{abstract}
The advent of affordable parallel computers such as Beowulf PC
clusters and, more recently, of  multi-core PCs has been highly
beneficial for  a large number of scientists and smaller institutions
that might not otherwise have access to substantial computing
facilities. However, there has not been an analogous progress in the
development and dissemination of parallel software: scientists need the
expertise to develop parallel codes and have to invest a significant
amount of time in the development of  tools even for the most common
data analysis tasks.
We  describe the Beowulf Analysis Symbolic INterface (BASIN) a
multi-user parallel data analysis and visualization framework.
BASIN is aimed at providing scientists with a suite of parallel
libraries for astrophysical data analysis along with general tools for
data distribution and parallel operations on distributed data to allow
them to easily develop new parallel libraries for their specific tasks.

\end{abstract}
\section{You have a Supercomputer.  Are you a\\Super-Programmer? }

The first ``Beowulf'' PC cluster [1] provided a proof
of concept that sparked a revolution in the use of off-the-shelf
computers for numerical simulations, data mining, and statistical
analysis of large datasets.  Because these clusters use commodity
components, this paradigm has provided an unbeatable ratio of
computing power per dollar, and over the past decade Beowulf-class
systems have enabled a wide range of high-performance applications on
department-scale budgets.

Beowulfs have unquestionably been highly beneficial in our own field
of Astrophysics, especially in smaller academic institutions that
might not otherwise have access to substantial computing facilities.
Beowulfs
have been successfully applied to numerical simulations that use a
variety of algorithmic approaches to study systems
spanning the range of scales from individual stars and supernovae
(see e.g. [2]) to the horizon size of the universe (see e.g.[3][4]). 

 Unfortunately, progress in the development of
affordable parallel computers has not been matched by analogous
progress in the development and dissemination of parallel software
aimed at providing scientists with the tools needed to take advantage
of the computing power of parallel machines.  Scientists with access
to a Beowulf cluster still need the expertise to develop parallel
codes, and have to spend a tremendous amount of time in the
development and testing of tools, even for the most common data
analysis tasks.  This is in stark contrast to the situation for serial
data analysis and simulations, for which a large number of standard
general-purpose (e.g. Matlab, Maple, R/Splus, Mathematica, IDL) and
specialized (e.g. IRAF for astronomical data analysis) tools and
libraries exist.

The recent advent of multi-core PCs has broadened still further the
base of commodity machines that enable computationally intensive
parallel simulations and data analysis of large datasets.  However,
this impressive technological advance serves also to make the lack of
general tools for parallel data analysis even more striking.  The
result is a significant barrier to entry for users or developers of
parallel computing applications.

As the gap between increasing parallel computing power and the
availability of software tools needed to exploit it has become more
and more evident, a number of projects have attempted to develop such
tools.  Our team has developed a package of parallel computational
tools---the Beowulf Analysis Symbolic INterface (\basinn)---to deal
with precisely these issues.  \basin is a suite of parallel
computational tools for the management, analysis and visualization of
large datasets.  It is designed with the idea that not all scientists
need to be specialists in numerics.  Rather, a user should be able to
interact with his or her data in an intuitive way.  In its current
form, the package can be used either as a set of library functions in
a monolithic \cpp program or interactively, from a \pyth shell, using
the \basin \pyth interface.

\basin is not the only package with this goal in mind.  This magazine
has recently presented descriptions of Star-P [5] (a commercial package
aimed at providing environments such as Matlab, Maple and other with
seamless parallel capabilities) and PyBSP [6] (a Python library for the
development of parallel programs following the bulk synchronous
parallel model).  As an open-source project, \basin growth is to be
driven by the needs and contributions of users and developers both in
the astrophysical community and, possibly, in other computationally
intensive disciplines.

\section{BASIN}

\begin{figure}[h]
\centerline{\includegraphics[height=5in,angle=-90]{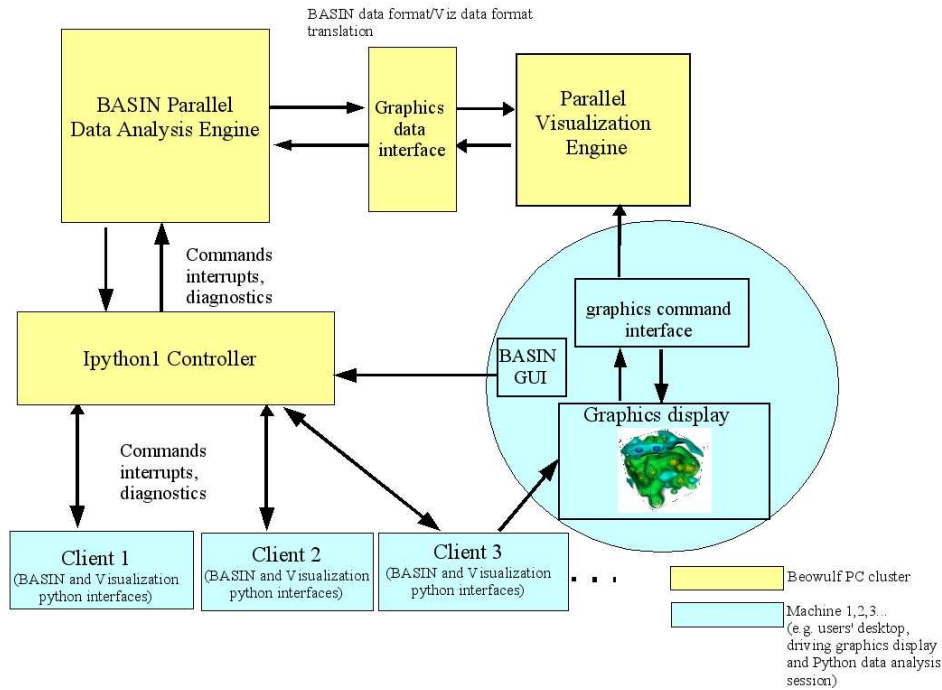}}
\caption{Schematic diagram illustrating the interaction between the
  main components of the \basin pipeline: the Data Analysis Engine,
  the User Interface, and the Graphics Display Engine. Several remote
  users (labeled in blue), may connect to and issue commands on an
  existing parallel compute engine running the BASIN kernel and
  visualization engine, and may share the same datasets in a
  collaborative data analysis session.}
\end{figure}

The three basic components of \basin are the Data Analysis Engine
(hereafter DAE), the User Interface (hereafter UI) and the Graphics
Display Engine (hereafter GDE).  Fig.~1 illustrates the relationships
among these system components.

\subsection*{The User Interface (UI)}

Although the entire set of \basin functions and tools can be accessed
from within a \cpp program using \basin as a library, the primary mode
of usage of \basin is one in which a user accesses its functionality
from within an interactive \pyth shell, applying the \basin analysis
and visualization tools to a ``live'' dataset during an interactive
data analysis session.  A Python GUI interface which provides
``point-and-click'' access to all the \basin functionality is also
available.

The communication between the client and the DAE is managed by the
tools included in
IPython1\footnote{http://ipython.scipy.org/moin/IPython1}. A
particularly important feature of the IPython1 package is the
capability of allowing multiple clients to connect to the same remote
engine.  Taking advantage of this feature, multiple distributed \basin
users can connect to the same DAE, share the same data and collaborate
on a data analysis session.  The UI includes functionality to keep
track of and save part or the entire history of a data analysis
session.  For a multi-user session, the history also includes
information on which of the participating users issued each command.

\subsection*{The Data Analysis Engine (DAE)}

The DAE is the core of the \basin package and is responsible for
parallel data management and all data analysis operations.  It is
implemented in C++/MPI-2 and \pyth wrappers are created using the
Boost.Python library.\footnote{http://www.boost.org}

The basic DAE objects visible to the user are {\regions} of data, the
{\data} objects associated with them and the individual \attrs
comprising these {\data} objects.  For example, for data coming from a
simulation, a \region might be a set of nested grids of density
fields, or arrays of particle properties.  On the other hand, for a
galaxy survey, a \region might represent a physical region of the sky,
or data taken over a particular period of time.

Two distinct types of distributed data are supported: \bgrid{s}
describe regularly sampled data, e.g. density values on a 3-D
Cartesian grid from a CFD calculation.  \blist{s} represent
irregularly sampled data, such as a set of arrays representing the
properties of particles in a snapshot from an $N$-body simulation, or
the right ascensions, declinations, and redshifts of galaxies in a
survey.  Each individual property in a \blist\ or \bgrid\ is stored in
an \attr which can be accessed by the user, and on which the user can
perform a variety of operations.  Some of the the main features of
\attrs are:

\begin{enumerate}

\item Physically distributed data stored in an \attr are accessed by
  the user as if they were a single data structure on a shared-memory
  machine.  The initial data distribution is automatically managed by
  the class constructor. For example,
\begin{verbatim} 
    >>> x = attribute_random(1000000,seed)
\end{verbatim}
  will generate $10^6$ random numbers uniformly distributed (by
  default) between 0 and 1, and automatically distribute them as
  evenly as possible across all nodes.  As in a shared-memory machine,
  global indexes can be used to access individual data elements. All
  the complexities of data distribution and retrieval in a
  distributed-memory cluster are handled internally and hidden to the
  user, who can operate as though working on a shared-memory
  environment.

\item \attrs can store $n$-dimensional arrays of both primitive
  datatypes and user-defined structures and objects.

\item All the math operators and the \c math library functions are
  overloaded to operate on \attrs in a parallel vectorized form. 
  For example, a new radius attribute may
  be computed from a list of cartesian components in a completely intuitive
  way:
\begin{verbatim} 
    >>> r = sqrt(pow(x,2)+pow(y,2))
\end{verbatim}
 
\item All relational and logical operators are similarly overloaded to
  operate in a parallel vectorized form on \attrs.

\item In order to take advantage of faster access to local data, the
  \attr class allows each process to locate its own block of data and,
  if possible, to work on just that block, thus avoiding unnecessary
  inter-process data transfer.

\end{enumerate}

We have developed libraries for several specific astrophysical
subfields.  For example, the {\tt StellarDynamics} library includes
many functions normally used in the analysis of $N$-body simulations
of galaxies and star clusters, such as the determination of the center
of the stellar distribution, measurement of characteristic scales
(such as the core, half-mass, tidal, and other Lagrangian radii),
robust estimates of local densities, and the computation of radial and
other profiles of physically important quantities, such as density,
velocity dispersion, ellipticity, anisotropy, mass function, and so
on.  Similarly, the {\tt Cosmology} library allows a user to perform
operations commonly encountered in cosmological contexts---for
example, read in a list of galaxy redshifts, select a cosmological
model, the compute a host of model-specific properties, such as
lookback time and comoving distance.  A user can transform between
observational parameters ($\alpha$, $\delta$, $z$), to Cartesian
coordinates, and quickly and easily view lengthy and/or complex
datasets.

\basin also contains more general libraries for important mathematical
operations, such as Fast Fourier Transforms and statistical analysis.
As often as possible, we have incorporated powerful existing packages
into the \basin framework.  For example, we have imported the
``Fastest Fourier Transform in the West'' (FFTW)\footnote{\tt
  http://fftw.org} package as our central FFT engine.  The \basin
internals handle all of the issues involved in packing and unpacking
data, and keeping track of whether the data are real or complex.  The
user simply invokes the transform via a one-line command.

Much of this functionality has been developed using the DAE tools for
data distribution and parallel operations, with minimal, or even no
explicit use of MPI.  Users without advanced knowledge of MPI can
therefore still add new \basin functions and create new libraries for
their own and general use.  For example, the \basin function for the
parallel calculation the center of mass of a stellar system [defined
as ${\bf r}_{cm}=\sum^N_{i=1}m_i {\bf r}_i/\sum^N_{i=1}m_i$ where
$m_i$ and ${\bf r}_i=(x_i,~y_i,~z_i)$ are the masses and coordinates of
the stars in the stellar system], might be implemented within \basin
as
\begin{verbatim}
vector<double> 
center_of_mass(Attribute &x, Attribute &y, Attribute &z, Attribute &m)
{
    vector<double> com(3);
    double m_sum=sum(m);
    com[0] = sum(m*x)/m_sum;
    com[1] = sum(m*y)/m_sum;
    com[2] = sum(m*z)/m_sum;
    return com;
}
\end{verbatim}
Data distribution, parallel execution of required operations, and
consolidation of the results of operations performed on individual
nodes can all be performed using \basin distributed data and parallel
analysis commands, without explicit use of MPI functions.  Thus users
can take advantage of the \basin libraries already developed and,
thanks to the low-level \basin tools, can in many cases expand \basin
as needed without being experts in parallel programming.

\subsection*{The Graphics Display Engine (GDE)}

The current implementation of \basin allows users to view data using
standard Python plotting libraries.  A number of functions interface
\basin with the popular
matplotlib\footnote{http://matplotlib.sourceforge.net} package and
with gnuplot.\footnote{http://www.gnuplot.info/} However, when
standard Python plotting tools are used, data are transfered to the local
client machine.  This  is done automatically by \basin without
requiring the user to issue any specific command to transfer the data
but while convenient for development purposes and
small  
datasets, this is clearly not a desirable approach for the very large
datasets \basin is intended to handle.

For the visualization of large distributed datasets, \basin
incorporates an interface to
VisIt [7],\footnote{https://wci.llnl.gov/codes/visit/} a parallel visualization
package developed at Lawrence Livermore National Laboratory.  As with
\basin commands, all VisIt visualization functions can be invoked from
a remote \pyth shell or from a GUI; \basin and VisIt can therefore
share the same \pyth client, from which it is possible to invoke both
\basin and VisIt commands which, in turn, are executed on the same
parallel computational engine (the \basin DAE).

\begin{figure}[h!]
\centerline{\includegraphics[height=5in,angle=-90]{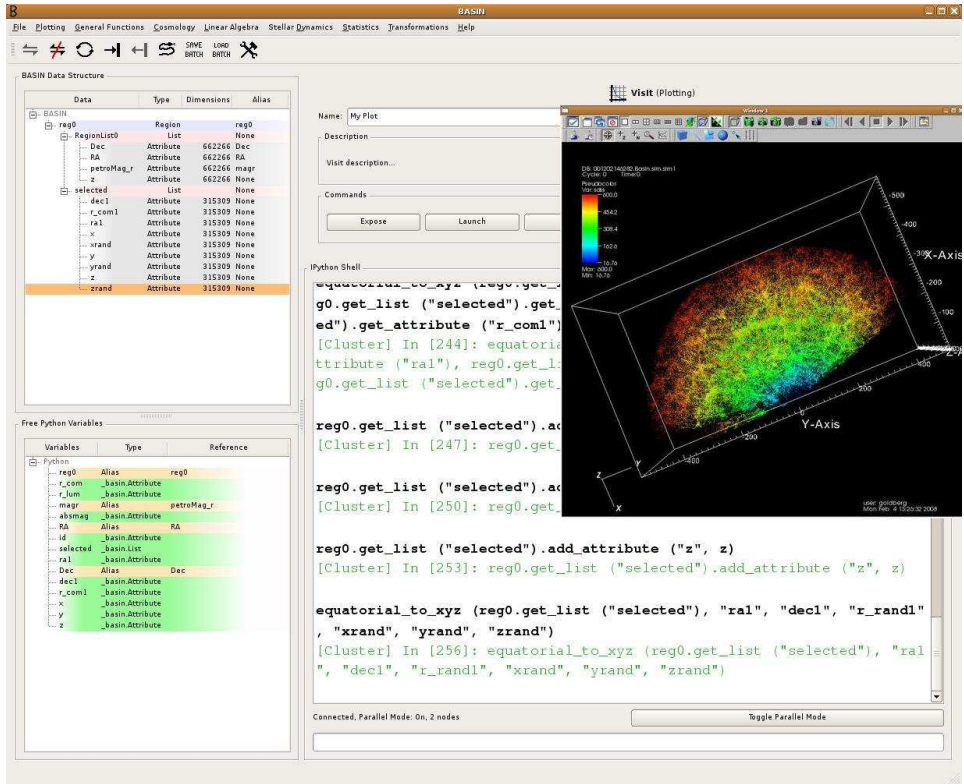}}
\caption{A BASIN session in which we read the SDSS DR6 spectroscopic
  sample, transform from observable redshift to cosmological
  distances, select a volume limited subsample, and plot the data with VisIt.}
\label{fg:sdss}
\end{figure}

As an example of a \basin analysis and visualization session, Figure
\ref{fg:sdss} shows a snapshot of a typical session within the \basin GUI,
which goes from a superset of the desired data to a selected,
transformed, and binned dataset.  The particular dataset in question
is the Sloan Digital Sky Survey (hereafter SDSS; [8]),
which contains over 10 Tbytes of ``high-quality'' reduced data,
including a sample of approximately 700,000 spectroscopically measured
galaxies in the sixth data release (DR6).

In this example, we can successfully (and simply,
from the user's perspective) analyze the SDSS spectroscopic galaxy
sample, performing coordinate transformations, producing a
volume-limited selection criterion, iteratively estimating the power
spectrum of the resulting density field, and visualizing the results,
all with 25 lines of scripting code.  The analysis is done entirely in
parallel, and serves as a template for how scientists, in an ideal
world, would like to interact with data.

The \basin GUI is also shown in Figure \ref{fg:sdss}.  In the upper
left, a user can drop and drag to rearrange a hierarchical dataset,
while the lower left shows a list of variables and data structures
available to the user.  The lower-right shows the equivalent Python
session, including commands which are automatically generated by
dragging and dropping data into various dialog boxes. 

From the \basin GUI, many of the commands of a data analysis session are
simple drag and drop operations, saving the user the effort even of
finding the correct syntax in the API.

\section{Examples}

The following examples illustrate how \basin allows a user to easily
develop scripts which transparently distribute data and perform
parallel operations.

\subsection*{Monte-Carlo calculation of $\pi$}
The Buffon-Laplace needle algorithm is a simple Monte Carlo algorithm
for the approximate calculation of $\pi$.  It has been used in
previous articles in this magazine ([5],[6]) to illustrate the use of
parallel software development tools.  
The algorithm is based on the estimation of the probability,
$P(l,a,b)$, that a needle of length $l$ thrown on a two-dimensional
grid with cells of length $a$ in one direction and $b$ in the other
direction will intersect at a least one line
$$
	P(l,a,b)={2l(a+b)-l^2 \over \pi a b}
$$
The ratio of the number of times the needle crosses the grid to the
total number of trials provides a numerical estimate of $P$ and allows
to calculate $\pi$.  
The implementation of this algorithm with a \basin \pyth
script is straightforward:

\begin{verbatim}
    >>> a = 1.0
    >>> b = 1.0
    >>> len = 0.6
    >>> numtrials = 10000000
    >>> phi = 2.*pi*attribute_random(numtrials)
    >>> x = a*attribute_random(numtrials)+len*cos(phi)
    >>> y = b*attribute_random(numtrials)+len*sin(phi)
    >>> count = sum(x<=0 | x>=a | y<=0 | y>=b)
    >>> pi = numtrials/float(count)*(2*len*(a+b)-len**2)/(a*b)
\end{verbatim}

\begin{figure}[h]

\centerline{\includegraphics[height=7cm]{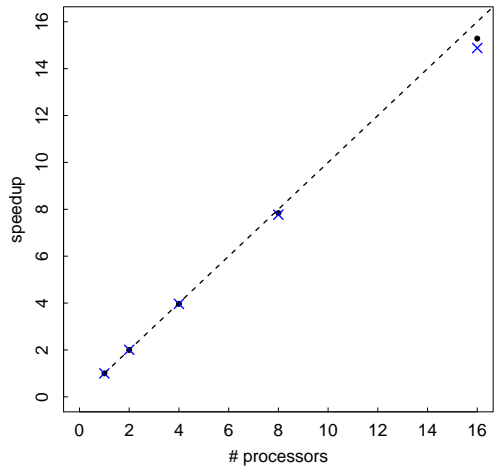}}
\caption{Scaling with the number of
  processors used for the Monte-Carlo calculation of $\pi$ with
  $N=10^7$ trials (filled dots) and for the direct force calculation
  in a $N$-body system with $N=5\times 10^6$ particles (crosses) discussed in section 
  3. The dashed line shows the ideal linear speed-up.  The
  benchmark was performed on 
  a Beowulf cluster with nodes each containing 2 AMD 1.5 Ghz chips.
  Nodes are connected via a channel-bonded 100 Mbit/s ethernet
  switch.} 
\label{fg:benchmark}
\end{figure}

 The speed-up of the
calculation for $N=10^7$ trials is shown in Figure~\ref{fg:benchmark}.


\subsection*{Direct force calculation in an $N$-body system}

$N$-body simulations are key computational tools for studying the
formation and evolution of stellar systems ranging from the relatively
small open and globular star clusters to much larger objects, such as
galaxies and galaxy clusters.

Depending on the type of stellar system and the time span of the
simulation, many sophisticated algorithms have been devised to
accelerate the calculation of the basic gravitational acceleration of
a star due to all other stars in the system.  However, the
approximations introduced to speed up the calculation are not always
adequate, and the computationally very expensive direct-force
calculation is in many cases still the only viable option.

The parallel calculation of the gravitational acceleration on a star
with position ${\bf r}_i=(x_i,y_i,z_i)$ due to all the other stars with mass
$m_j$ and position ${\bf r}_j$ in a stellar system (including softening)
$$
	{\bf a}=\sum_{j} G m_j {{\bf r}_j-{\bf r}_i
			\over (\epsilon^2 +|{\bf r}_i-{\bf r}_j|^2)^{3/2}}\,.
$$
is straightforward in \basin/Python, as illustrated by the following
code fragment:
\begin{verbatim}
>>> xi = x[i]
>>> yi = y[i]
>>> zi = z[i]
>>> d = G*m/pow(eps2+(pow(x-xi,2)+pow(y-yi,2)+pow(z-zi,2)),1.5)
>>> ax = sum((x-xi)*d)
>>> ay = sum((y-yi)*d)
>>> az = sum((z-zi)*d)
\end{verbatim}
Each process calculates the contribution to the acceleration due only
to stars stored locally.  Access to the coordinates of the star on
which the acceleration is being calculated is straightforward, and the
user does not need to perform any complex operation, or explicitly issue any
MPI command, in order to locate the node on which the coordinates are
stored or broadcast that information to other nodes.  Any element of
an \attr can be accessed through the bracket operator as on a shared
memory machine.  Data distribution and remote access, parallel
vectorized operations, reduction operations and interprocess
communication, are all handled by the \basin tools and are completely
hidden to the user.

\section{Looking to the Future}

\basin is still under active development, and will be expanded in
numerous ways in the near future.  Currently, the main areas of
development are:
\begin{itemize}

\item Benchmarking and scalability. We are currently studying the
  performance and the scalability of \basin on the NSF Teragrid
  superclusters with thousands of processors.

\item \basin for Graphics Processing Units (GPUs).  General programming on GPUs
  is becoming increasingly popular, in large part due to the Nvidia
  CUDA language, which provides programmers with a tool to more easily
  harness the enormous computational power of GPUs.  We are currently
  in the process of porting all atomic BASIN operations to the GPU
  architecture.  The machine we envisage for the GPU-enabled \basin is
  a Beowulf cluster with each node hosting one or more GPUs.  Data
  will still be distributed over all the nodes of a cluster, but the
  atomic operations performed internally by each node will be carried
  out on, or accelerated by, the GPU.  The programming model therefore
  has two level of parallelism: the standard cluster-level
  parallelism, with data distributed by the MPI-based \basin tools,
  and internal parallelism on each node, in which the capabilities of
  GPUs are used to further accelerate assigned operations.

\item Beyond MPI.  All parallel functions and distributed data
  structures in \basin are implemented using MPI.  However, new
  languages and extensions to existing languages are currently being
  developed to facilitate the development of parallel codes.
  Specifically, Partitioned Global Address Space languages (see
  e.g. [9])  such as co-array Fortran, Unified Parallel C
  (UPC), and Titanium seem to provide interesting alternatives to
  MPI-based code.  We plan to build a UPC-based implementation of all
  the low-level \basin distributed data structures, to compare its
  performance and scalability with the MPI version.

\item Beyond Astrophysics. Although the \basin libraries currently available
are focussed on the analysis of astrophysical data, all the \basin
tools for parallel operations and data distribution are completely
general and can be used to develop packages relevant to other
disciplines.  We plan to expand \basin and broaden its capabilities by 
collaborating with scientists in other fields, such computational
biology and information sciences, involving computationally intensive
data analysis and simulations.  

\item Integration with other science packages.  We plan to integrate
  \basin with the MUltiscale MUltiphysics Simulation Environment
  (MUSE),\footnote{http://muse.li} a framework for the numerical study
  of the multi-scale physics of dense stellar systems.  The result
  will be a single environment in which a user can run novel numerical
  simulations, including a wide variety of physical processes, and,
  within the same platform, perform parallel analysis and
  visualization of the results, in real time or post-production.

\end{itemize}
  
\section*{Acknowledgments}
We thank all the members of the BASIN team (B. Char, D. Cox, A. Dyszel,
J. Haaga, M. Hall,  L. Kratz, S. Levy, P. MacNeice, E. Mamikonyan,
M. Soloff, A. Tyler, M. Vogeley and B. Whitlock) for many discussions and
comments on the issues presented in this paper.  

\section*{References}

[1] Becker, D.J., Sterling, T.J., Savarese, D.F., Dorband, J.E.,
  Ranawak, U.A., \& Packer, C.V., 1995, Proceedings of the
  International Conference on Parallel Processing\\

[2] Blondin, J.M., ``Discovering new dynamics of core-collapse supernova
  shock waves'', 2005, Journal of Physics: Conference Series 16 370 \\

[3] Norman, M.L., Bryan, G.L., Harkness, R., Bordner, J., 
Reynolds, D., O'Shea, B. \& Wagner, R., `` Simulating Cosmological
Evolution with Enzo'' To appear in Petascale Computing: Algorithms and
Applications, Ed. D. Bader, CRC Press LLC (2007), arXiv:0705.1556v1\\

[4] Springel V. et al. 2005, ``Simulations of the formation, evolution and clustering of galaxies and quasars'', Nature 435 629\\

[5] S. Raghunathan, ``Making a Supercomputer Do What You Want: High-Level
Tools for Parallel Programming'', {\it Computing in Science and
  Engineering}, vol. 8, no. 5, 2006, pp.70-80\\

[6] K. Hinsen, ``Parallel Scripting with Python'', {\it Computing in
  Science and Engineering}, vol. 9, no. 6, 2007, pp.82-89\\

[7] Childs H., Brugger E., Bonnell K., Meredith J., Miller M.,
Whitlock B., Max N.: A contract based system for large data
visualization. In Proc. of Visualization 2005 Conference. 191\\

[8] D. York, et al., ``The Sloan Digital Sky Survey: Technical Summary'',
{\it Astronomical Journal}, vol. 120, 2000, p. 1579\\

[9] Carlson, B., El-Ghazawi, T., Numrich, R., Yelick K.,  2003, ``Programming in the Partitioned Global Assdress Space Model'', https://upc-wiki.lbl.gov/UPC/images/b/b5/PGAS
\_ Tutorial\_ sc2003.pdf\\

\end{document}